\documentclass[13pt]{article}
\usepackage{mathrsfs}
\usepackage{amsfonts}
\usepackage{amsmath,amssymb}
\begin{document}
\title{{\bf On Fixed Points of L\"{u}ders Operation}\thanks{This project is supported by Natural Science Found of China (10771191 and
10471124).}}
\author {{Liu Weihua,\,\,  Wu Junde\thanks{Corresponding author E-mail: wjd@zju.edu.cn}}\\
{\small\it Department of Mathematics, Zhejiang University, Hangzhou
310027, P. R. China}}

\date{}
\maketitle

{\small\it \noindent {\bf Abstract}. In this paper, we give a
concrete example of a L\"{u}ders operation $L_{{\cal A}}$ with $n=3$
such that $L_{{\cal A}}(B)=B$ does not imply that $B$ commutes with
all $E_1, E_2$ and $E_3$ in $\cal A$, this example answers an open
problem of Professor Gudder.

\vskip 0.1 in

}.

\vskip 0.2 in

\noindent Let $H$ be a complex Hilbert space, ${\cal B}(H)$ be the
bounded linear operator set on $H$, ${\cal E}(H)=\{A: 0\leq A\leq
I\}$, ${\cal A}=\{E_i\}_{i=1}^n\subseteq {\cal E}(H)$ and
$\sum_{i=1}^nE_i=I$, where $1\leq n\leq\infty$. The famous {\it
L\"{u}ders operation} $L_{{\cal A}}$ is a map which is defined on
${\cal B}(H)$ by:

$$L_{{\cal A}}: A\rightarrow \sum\limits_{i=1}^{n}E_{i}^{\frac{1}{2}}A
E_{i}^{\frac{1}{2}}.$$

A question related to a celebrated theorem of L\"{u}ders operation
is whether $L_{{\cal A}}= A$ for some $A\in {\cal E}(H)$ implies
that $A$ commutes with all $E_i$ for $i=1, 2, \cdots, n$ ([1]). The
answer to this question is positive for $n=2$ ([2]), and negative
for $n=5$ ([1]). In this paper it is shown, by using a simple
derivation of the example of Arias-Gheondea-Gudder in [1], that the
answer is negative as well for $n=3$, a question raised by Gudder in
2005 ([3]).

First,  we denote ${\cal B}(H)^{L_{{\cal A}}}=\{B\in {\cal B}(H):
L_{{\cal A}}(B)=B\}$ is the fixed point set of $L_{{\cal A}}$,
${\cal A}'$ is the commutant of ${\cal A}$.

\vskip 0.1 in

{\bf Lemma 1} ([1]). If ${\cal B}(H)^{L_{{\cal A}}}={\cal A}'$, then
${\cal A}'$ is injective.

\vskip 0.1 in

{\bf Lemma 2} ([1]). Let $F_2$ be the free group generated by two
generators $g_1$ and $g_2$ with identity $e$, $\mathbb{C}$ be the
complex numbers set and $H=l_2(F_2)$ be the separable complex
Hilbert space
$$H=l_2(F_2)=\{f|f:F_2\rightarrow \mathbb{C}, \sum
|f(x)|^2<\infty\}.$$ For $x\in F_2$ define $\delta_x: F_2\rightarrow
C$ by $\delta_x(y)$ equals $0$ for all $y\neq x$ and $1$ when $y=x$.
Then $\{\delta_x|x\in F_2\}$ is an orthonormal basis for $H$. Define
the unitary operators $U_1$ and $U_2$ on $H$ by
$U_1(\delta_x)=\delta_{g_1x}$ and $U_2(\delta_x)=\delta_{g_2x}$.
Then the von Neumann algebra $\mathscr{N}$ which is generated by
$U_1$ and $U_2$ and its commutant $\mathscr{N}'$ are not injective.

\vskip 0.1 in

Now, we follow the Lemma 1 and Lemma 2 to prove our main result:

\vskip 0.1 in

Let $\mathbb{C}_1$ be the unite circle in $\mathbb{C}$ and $h$ be a
Borel function be defined on the $\mathbb{C}_1$ as following:
$h(e^{i\theta})=\theta$ for $\theta\in [0,2\pi)$. Then $A_1=h(U_1)$
and $A_2=h(U_2)$ are two positive operators in $\mathcal {N}$. If
take the real and imagine parts of $U_1=V_1+iV_2$ and
$U_2=V_3+iV_4$, then $\mathcal {N}$ is generated by the self-adjoint
operators $\{V_1,V_2,V_3,V_4\}$ ([1]). Since functions $\cos $ and
$\sin $ are two Borel functions, so we have
$V_1=\frac{1}{2}(U_1+U_1^*)=\cos(A_1)$, $V_2=\sin(A_1)$,
$V_3=\cos(A_2)$ and $V_4=\sin(A_2)$. Thus $\mathscr{N}$ is contained
in the von Neumann algebra which is generated by $A_1$ and $A_2$.

On the other hand, it is clear that the von Neumann algebra which is
generated by $A_1$ and $A_2$ is contained in $\mathscr{N}$. So
$\mathscr{N}$ is the von Neumann algebra which is generated by $A_1$
and $A_2$. Let $E_1=\frac{A_1}{2\|A_1\|}$,
$E_2=\frac{A_2}{2\|A_2\|}$ and $E_3=I-E_1-E_2$. Then ${\cal
A}=\{E_1,E_2,E_3\}\subseteq {\cal E}(H)$ and $E_1+E_2+E_3=I$.

Now, we define the L\"{u}ders operation on ${\cal B}(H)$ by
$$L_{{\cal A}}(B)=\sum\limits_{i=1}^{3} E_i^{1/2}BE_i^{1/2}.$$ It is clear that the Von Neumann algebra which is generated by $\{E_1, E_2, E_3\}$ is
$\mathscr{N}$, so it follows from Lemma 1 and Lemma 2 that
$B(H)^{L_{{\cal A}}}\supsetneq{\cal A}'$, thus there exists a $D\in
B(H)^{L_{{\cal A}}}\setminus{\cal A}'$. Now, the real part or the
imaginary part $D_1$ of $D$ also satisfies $D_1\in B(H)^{L_{{\cal
A}}}\setminus {\cal A}'$. Let $D_2=||D_1||I-D_1$. Then $D_2\geq 0$.
Let $D_3=\frac{D_2}{||D_2||}$. Then $D_3\in {\cal E}(H)$ and $D_3\in
B(H)^{L_{{\cal A}}}\setminus{\cal A}'$. Thus, we proved the
following theorem which answered the question in [3].

{\bf Theorem 1}. Let $H=l_2(F_2)$, ${\cal A}=\{E_i\}_{i=1}^3$ be
defined as above. Then there is a $B\in {\cal E}(H)$ such that
$L_{{\cal A}}(B)=B$, but $B$ does not commute with all $E_1, E_2$
and $E_3$.

\vskip0.2in

{\bf Acknowledgement.} The authors wish to express their thanks to
the referee for his (her) important comments and suggestions.

\vskip0.3in

\centerline{\bf References}

\vskip0.2in

\noindent [1] A. Arias, A. Gheondea, S. Gudder. Fixed points of
quantum operations. J. Math. Phys., 43, 2002, 5872-5881

\noindent [2] P. Busch, J. Singh. L\"{u}ders theorem for unsharp
quantum measurements. Phys. Letter A, 249, 1998, 10-12

\noindent [3] S. Gudder. Open problems for sequential effect
algebras. Inter. J. Theory. Physi. 44, 2005, 2199-2205

\end{document}